\documentclass{ws-ijmpd}
\usepackage{graphicx}
\usepackage{subfigure}
\begin{document}

\markboth{Wei-Zhong Liu ,  Dao-Jun Liu} {Statefinder Diagnostic for
Quintessence with or without Thermal Interaction}

%
\catchline{}{}{}{}{}
%

\title{STATEFINDER DIAGNOSTIC FOR QUINTESSENCE WITH OR WITHOUT THERMAL INTERACTION  }

\author{Wei-Zhong Liu}

\address{Center for Astrophysics, Shanghai Normal University, 100 Guilin
Road\\
  Shanghai, 200234, China}

\author{Dao-Jun Liu}

\address{Center for Astrophysics, Shanghai Normal University, 100 Guilin
Road\\
  Shanghai, 200234, China\\
djliu@shnu.edu.cn}

\maketitle


\begin{abstract}
The cosmological dynamics of minimally coupled scalar field that
couples to the background matter with thermal interactions is
investigated by using statefinder diagnostics. The time evolution of
the statefinder pairs $\{r,s\}$ and $\{r,q\}$ are obtained under the
circumstance that different values of model parameters are chosen.
We show that the thermal coupling term does not affect the location
of the late-time attractor, but exert an influence on the evolution
of the statefinder parameters. The most notable feature of the $r-s$
plane for the thermal coupling model which is distinguished from the
other dark energy models is that some part of the curve with thermal
coupling can form a closed loop in the second quadrant ($r>1,s<0$).
\end{abstract}


\section{Introduction}
Various of cosmological observations strongly suggest that our
universe is undergoing an accelerated expansion phase. In order to
explain the acceleration, an unexpected  energy component of the
cosmic budget, which dominate the universe only recently,  is
introduced by many cosmologists. However, the physical origin of
dark energy is still mysterious.

Perhaps the simplest proposal is the Einstein's cosmological
constant $\Lambda$ (vacuum energy), whose energy density remains
constant with time. However, due to some conceptual problems
associated with the cosmological constant (for a review, see
\cite{ccp,ccp2,ccp3}), a large variety of alternative possibilities
have been explored. The most popular among them is quintessence
scenario which uses a scalar field $\phi$ with a suitably chosen
potential $V(\phi)$ so as to make the vacuum energy vary with time.
A form of quintessence called "tracker fields", whose evolution is
largely insensitive to initial conditions and at late times begin to
dominate the universe with a negative equation of state, was
introduced to avoids the problems relevant to cosmological constant
\cite{Zlatev}. Another approach to solve the puzzle is introducing
an interaction term in the equations of motion, which describes the
energy flow between the dark energy and the rest matter (mainly the
dark matter) in the universe. It is found that, with the help of a
suitable coupling, it is possible to reproduce any scaling
solutions. Inclusion of a non-minimal coupling to gravity in
quintessence models together with further generalization leads to
models of dark energy in a scalar-tensor theory of gravity. Besides,
some other models invoke unusual material in the universe such as
Chaplygin gas, tachyon, phantom or k-essence (see, for a review,
\cite{dde} and reference therein).

As so many dark energy models have been proposed, it becomes urgent
to give them an unambiguous discrimination. The equation of state
$w$, could discriminate some basic dark energy models properly, for
example, $w=-1$ for cosmological constant, $w>-1$ for quintessence
and $w<-1$ for phantom, but as more and more interacting models were
investigated, the ambit of $w$ become not so clear. A new
geometrical diagnostic, dubbed the statefinder pair $\{r, s\}$ is
proposed by Sahni \textit{et al} \cite{statefinder}, where $r$ is
only determined by the scalar factor $a$ and its derivatives with
respect to the cosmic time $t$, just as the Hubble parameter $H$ and
the deceleration parameter $q$, and $s$ is a simple combination of
$r$ and $q$. The statefinder pair has been used to explor a series
of dark energy and cosmological models
\cite{SF03,xzhang05a,xzhang05b,sf}. As is analyzed, the "distance"
from a given dark energy model to the LCDM scenario can be clearly
identified via the $r(s)$ evolution diagram. The current values of
$r$ and $s$, which can be calculated in models, are evidently
valuable since it is expected that they can be extracted from data
coming from SNAP (SuperNovae Acceleration Probe) type experiments.
Therefore, the statefinder diagnostic combined with future
observations may possibly be used to discriminate between different
dark energy models \cite{xzhang05a,xzhang05b}.

The effects of thermal coupling between the quintessence field and
the ordinary matter particles was first investigated by Hsu and
Murray \cite{Hsu}.
 They set the quintessence field to be a static external
source for a Euclidean path integral depicting the thermal degree of
freedom and let the time-like boundary conditions of the path
integral have a period which is decided by the inverse of
temperature. And they show that if in the early universe matter
particles are in thermal equilibrium, quantum gravity
\cite{Kamionkowski} will induce an effective thermal mass term for
quintessence field $\phi$, which takes the form
\begin{equation}\label{eq1}
\left(\frac{\beta}{M_P}\right)^2\phi^2 T^4,
\end{equation}
where $T\propto a^{-1}$ denotes a measurement of temperature of
background matter, $M_P$ the Planck mass scale, $\beta$ a
dimensionless constant, $a$ the scale factor (with current value
$a_0=1$), and find that even Planck-suppressed interactions between
matter and the quintessence field can alter its evolution
qualitatively. In a previous paper \cite{lls}, we investigated the
dynamics of the cosmology with quintessence using the phase space
analysis, which has the above thermal coupling to the matter in a
complete manner and analyzed  the conditions for the existence and
stability of various critical points as well as their cosmological
implications. In this paper, we apply the statefinder diagnostic to
the cosmological dynamics of minimally coupled scalar field that
couples to the background matter with  thermal interactions.

\section{The Setup}
Let us study  quintessence with  thermal coupling to ordinary matter
particles in spatially flat FRW cosmological background
\begin{equation}\label{metric}
ds^2=dt^2-a^2(t)d\textbf{x}^2.
\end{equation}
For the spatially homogeneous scalar field minimally coupled to
gravity with thermal interaction (\ref{eq1}), the evolution is
governed by the Klein-Gordon equation
\begin{equation}\label{ddphi}
\ddot{\phi}+3H\dot{\phi}+V'(\phi)=2\left(\frac{\beta}{M_P}\right)^2\phi
T^4,
\end{equation}
where the overdots denote the derivative with respect to cosmic time
and the prime denotes the derivative with respect to $\phi$. Here
the Hubble parameter $H\equiv \dot{a}/a$ is determined by the
Friedmann equation
\begin{equation}\label{FRW}
H^2=\frac{\kappa^2}{3}\left[\rho_m+\frac{1}{2}\dot{\phi}^2+V(\phi)\right]
\end{equation}
and
\begin{equation}\label{dH}
\dot{H}=-\frac{\kappa^2}{2}(\rho_m+p_m+\dot{\phi}^2),
\end{equation}
where $\kappa^2\equiv 8\pi/M_p^2$, $\rho_m$ and $p_m$ are the energy
density and pressure of the baryotropic matter, respectively. From
Eqs.(\ref{ddphi})-(\ref{dH}) and the conservation of energy,
$\rho_m$ satisfies the following continuous equation
\begin{equation}\label{sys1}
\dot{\rho_m}+3H(\rho_m+p_m)=-2\left(\frac{\beta}{M_P}\right)^2\phi\dot{\phi}T^4,
\end{equation}
 and $p_m=(\gamma-1)\rho_m$, where $\gamma$ is a
constant, $0\leq \gamma\leq2$, such as radiation($\gamma=4/3$) or
dust ($\gamma=1$). It is clear that when the thermal coupling
parameter $\beta$ becomes zero, the equations
(\ref{ddphi})-(\ref{sys1}) will return to those of the standard one
scalar field quintessence scenario. In current situation, both
quintessence and baryotropic matter are not conserved, but the
conservation of the overall energy holds.

To be concrete, in this paper we only consider the quintessence with
an exponential potential energy density, \textit{i.e.},
\begin{equation} \label{potential1}
 V(\phi)=V_0\exp(-\lambda_0 \kappa \phi)
 \end{equation}
 where the parameter $V_0$ and $\lambda_0$ are two positive constants.
 Exponential potentials have been studied extensively in various situations, and these are of interest for two main reasons. Firstly, they can be derived from a good candidate of fundamental theory for such being string/M theory; secondly, the equations of motion can be written as an autonomous system in the situation.

 Introducing  the following dimensionless variables:
\begin{eqnarray}\label{defx}
x = \frac{\kappa\dot{\phi}}{\sqrt{6}H}, \;\;y =
\frac{\kappa\sqrt{V(\phi)}}{\sqrt{3}H}, \;\;z =
\frac{\kappa}{M_p}\frac{\phi}{\sqrt{3}a^2H}, \;\;\xi  =
\frac{\sqrt{6}}{\kappa \phi}, \;\;N = \ln a
\end{eqnarray}
and taking the potential (\ref{potential1}) into account,  we can
rewrite the equation system (\ref{FRW})-(\ref{sys1}) as the
following autonomous
 system:

\begin{eqnarray}\label{auto1}
\frac{dx}{dN}&=&\frac{3}{2}x[\gamma(1-x^2-y^2)+2x^2]-(3x+\beta^2z^2\xi-\frac{1}{2}\lambda_0
y^2),\nonumber\\
\frac{dy}{dN}&=&\frac{3}{2}y[\gamma(1-x^2-y^2)+2x^2]-\frac{1}{2}\lambda_0
xy,\nonumber\\
\frac{dz}{dN}&=&\frac{3}{2}z[\gamma(1-x^2-y^2)+2x^2]-2z+xz\xi,\nonumber\\
\frac{d\xi}{dN}&=&-x\xi^2.
\end{eqnarray}

\section{Statefinder Parameters}

The traditional geometrical diagnostics, \textit{i.e.}, the Hubble
parameter $H$ and the deceleration parameter $q\equiv
{-\ddot{a}a}/{\dot{a}^2}$, are two good choices to describe the
expansion state of our universe but they can not characterize the
cosmological models uniquely, because a quite number of models may
just correspond to the same current value of $H$ and $q$.
Fortunately, as is shown in many literatures, the statefinder pair
$\{r, s\}$ which is also a geometrical diagnostic, is able to
distinguish a series of cosmological models successfully.

The statefinder pair $\{r,s\}$ defines two new cosmological
parameters in addition to $H$ and $q$:
\begin{equation}\label{defr}
r\equiv \frac{1}{aH^3}\frac{d^3a}{dt^3},\;\;\;\; s\equiv
\frac{r-1}{3(q-1/2)}.
\end{equation}

As an important function, the statefinder can allow us to
differentiate between a given dark energy model and the simplest of
all models,  i.e., the cosmological constant $\Lambda$. For the
$\Lambda$CDM model, the statefinder diagnostic pair $\{r,s\}$ takes
the constant value $\{1,0\}$, and for the SCDM model, $\{1,1\}$.

 From Eqs. (\ref{defr}) and (\ref{defx}), the statefinder parameter $r$ can be explicitly written as
\begin{eqnarray}\label{exprs}
r=1+\frac{9}{2}\gamma(\gamma-1)(1-x^2-y^2)
+3(\gamma-2)\beta^2xz^2\xi+9x^2-3\lambda_0 xy.
\end{eqnarray}
And the deceleration parameter $q$ is given by
\begin{equation}\label{expq}
q=\frac{3}{2}\gamma(1-x^2-y^2)+3x^2-1.
\end{equation}
Therefore, the other statefinder parameter $s$ is equal to
\begin{equation}\label{s}
    s=\frac{\gamma(\gamma-1)(1-x^2-y^2)+\frac{2}{3}(\gamma-2)\beta^2xz^2\xi+2x^2-\frac{2}{3}\lambda_0xy}{\gamma(1-x^2-y^2)+2x^2-1}.
\end{equation}

\begin{figure}
\begin{center}
\includegraphics[height=6.cm]{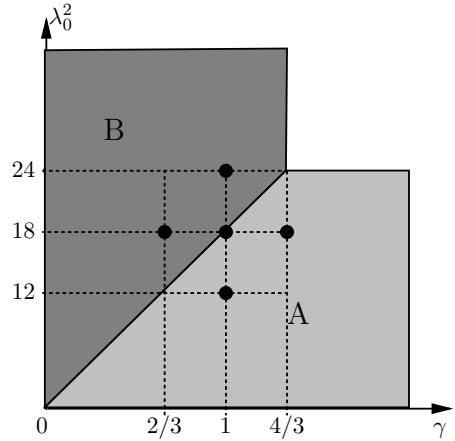}
\caption{The schematic illustration of the parameter space. In
region A and B, the autonomous system (\ref{auto1}) has a late-time
attractor, respectively.  }\label{para}
\end{center}
\end{figure}

We draw a $\gamma-\lambda^2$ plane in Figure \ref{para} and divide
it into different regions. From the  analysis of our previous paper
\cite{lls}, we know that the equation system, Eqs.(\ref{auto1}),
have a sole late-time attractor only if  the conditions $\gamma<4/3$
and/or $\lambda_0^2<24$ are satisfied. It is easy to find that these
conditions correspond to region A and B in Figure \ref{para} and the
common boundary of region A and B  is the segment of straight line
$\lambda_0^2=18\gamma$. In what follows, we call the condition that
region A satisfies case A and that region B satisfies case B.  For
case A and case B, the attractor $(x_c,y_c,z_c,\xi_c)$ is located at
the point $(\lambda_0/6,\sqrt{36-\lambda_0^2}/6,0,0)$ and
$(3\gamma/\lambda_0,3\sqrt{\gamma(2-\gamma)}/\lambda_0,0,0)$ in the
phase space, respectively. It is readily to check that in the common
boundary of region A and B, there actually exists only one attractor
as is expected. It is also worth pointing out that the attractor
depends only on the parameter $\lambda$ for case A, and for both
cases, the attractor has nothing to do with the coupling parameter
$\beta$. In statefinder parameter space, the attractor lies at the
point $(r_c,q_c,s_c)$ and it can be expressed explicitly by
\begin{eqnarray}\label{attractor1}
r_c&=&1+\frac{\lambda_0^2}{12}\left(3-\sqrt{36-\lambda_0^2}\right),\;\;\nonumber\\
q_c&=&-1+\frac{\lambda_0^2}{12},\;\;\nonumber\\
s_c&=&\frac{{\lambda_0^2}\left(3-\sqrt{36-\lambda_0^2}\right)}{3\left(\lambda_0^2-18\right)}
\end{eqnarray}
for case A and
\begin{eqnarray}\label{attractor2}
r_c&=&1+\frac{9}{2}\gamma(\gamma-1) +\frac{81 (\gamma-2) \gamma
^2}{\lambda _0^2}
-\frac{27 \gamma  \sqrt{(2-\gamma ) \gamma }}{\lambda _0},\nonumber\\
q_c&=&-1+\frac{3 \gamma }{2},\nonumber\\
s_c&=&\gamma+\frac{18 (\gamma-2) \gamma^2}{(1-\gamma ) \lambda
_0^2}+\frac{6\gamma \sqrt{(2-\gamma ) \gamma }}{(1-\gamma ) \lambda
_0}
\end{eqnarray}
for case B, respectively.

Let us now turn to statefinder parameters plane analysis. We shall
investigate the properties of the evolution trajectories of
statefinder parameters with three parameters $\beta$, $\gamma$ and
$\lambda_0$, by fixing two of them and varying the rest. Note that
we are only interested in case A and case B, due to the existence of
the attractor solution.

\begin{figure}
\begin{center}
\mbox{\subfigure{\includegraphics[width=0.48\textwidth]{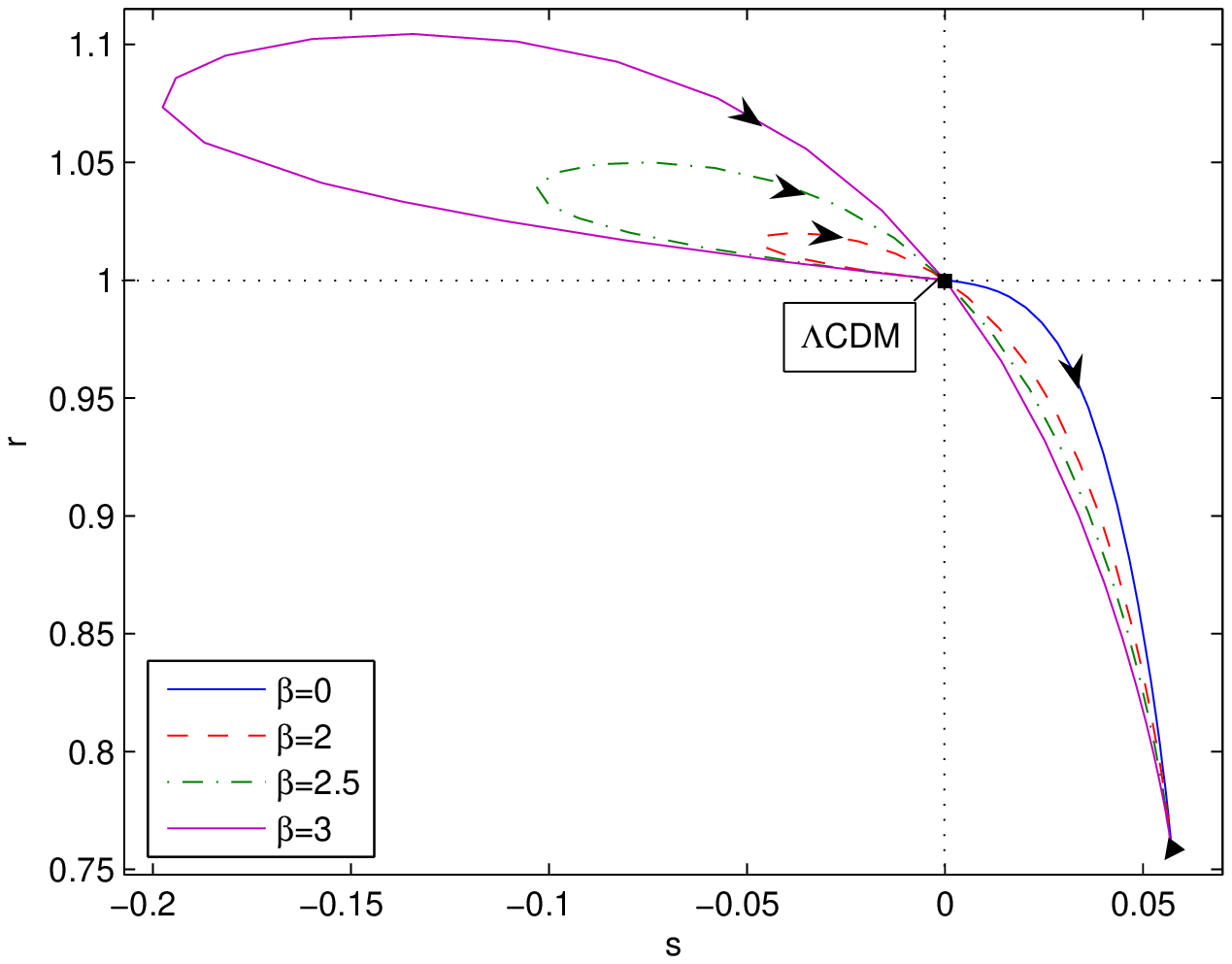}}
      \subfigure{\includegraphics[width=0.48\textwidth]{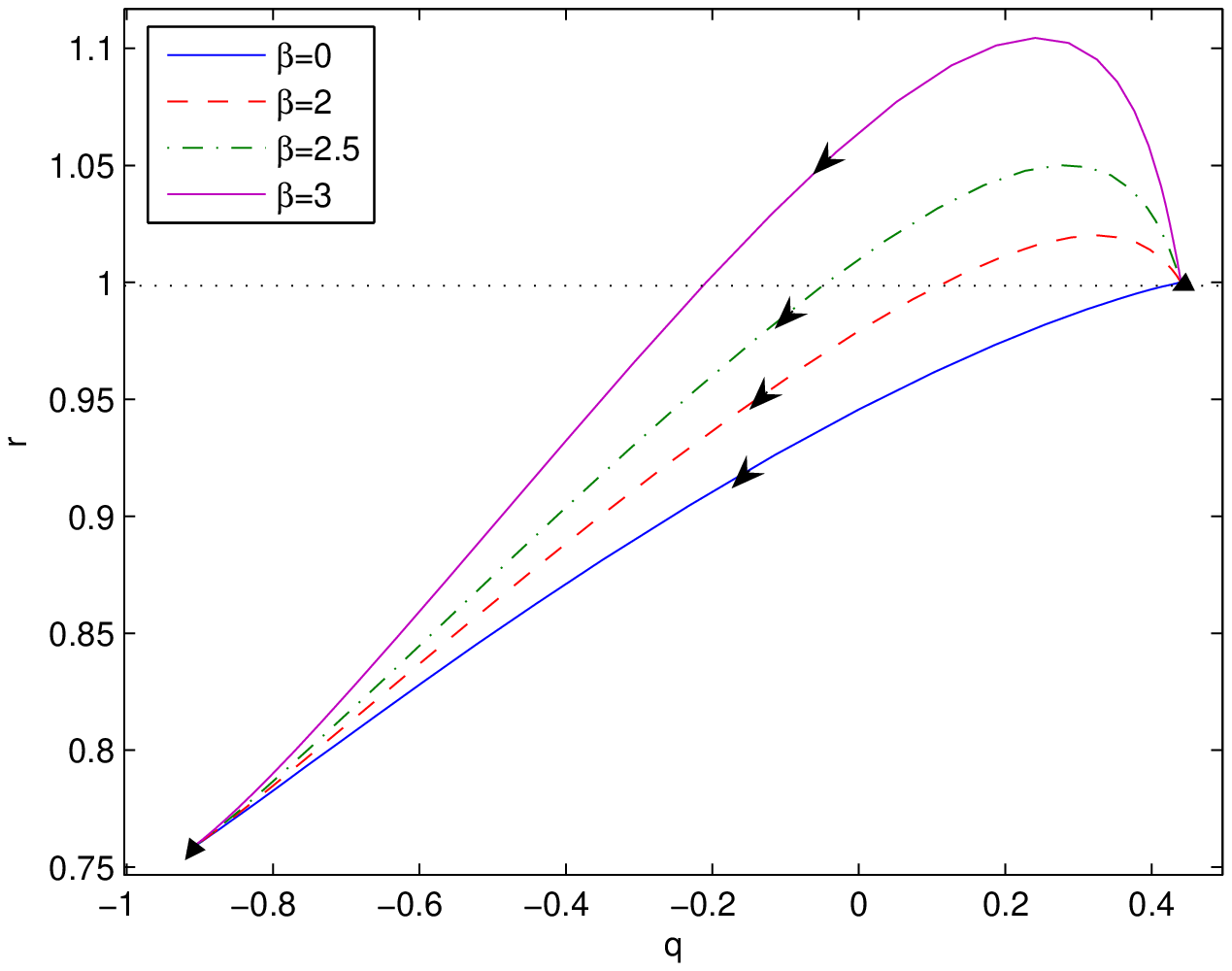}}}
\caption{The time evolution of the statefinder pairs $\{r,s\}$ and
$\{r,q\}$ is illustrated in left and right panel, respectively. In
both panels the model parameter $\gamma$ and $\lambda_0$ are fixed
to be unity and the thermal coupling constant $\beta$ takes
different values. For $\beta=0$ curve, $r$ and $s$ monotonically
decrease from $1$ and  increase from $0$ to the value the attractor
takes, respectively, and $q$ also monotonically decrease from
positive to negative. For $\beta\neq 0$ curves, $q$ keeps
monotonically decreasing, whereas $r$ ($s$) first increase
(decrease) from $1$ ($0$)  to a maximum (minimum) value then
decrease (increase) and pass through the $\Lambda$CDM fixed point
(in $r-s$ plane) to the attractor. }\label{fig:l1g1}
\end{center}
\end{figure}

\begin{figure}
\begin{center}
\mbox{\subfigure{\includegraphics[width=0.48\textwidth]{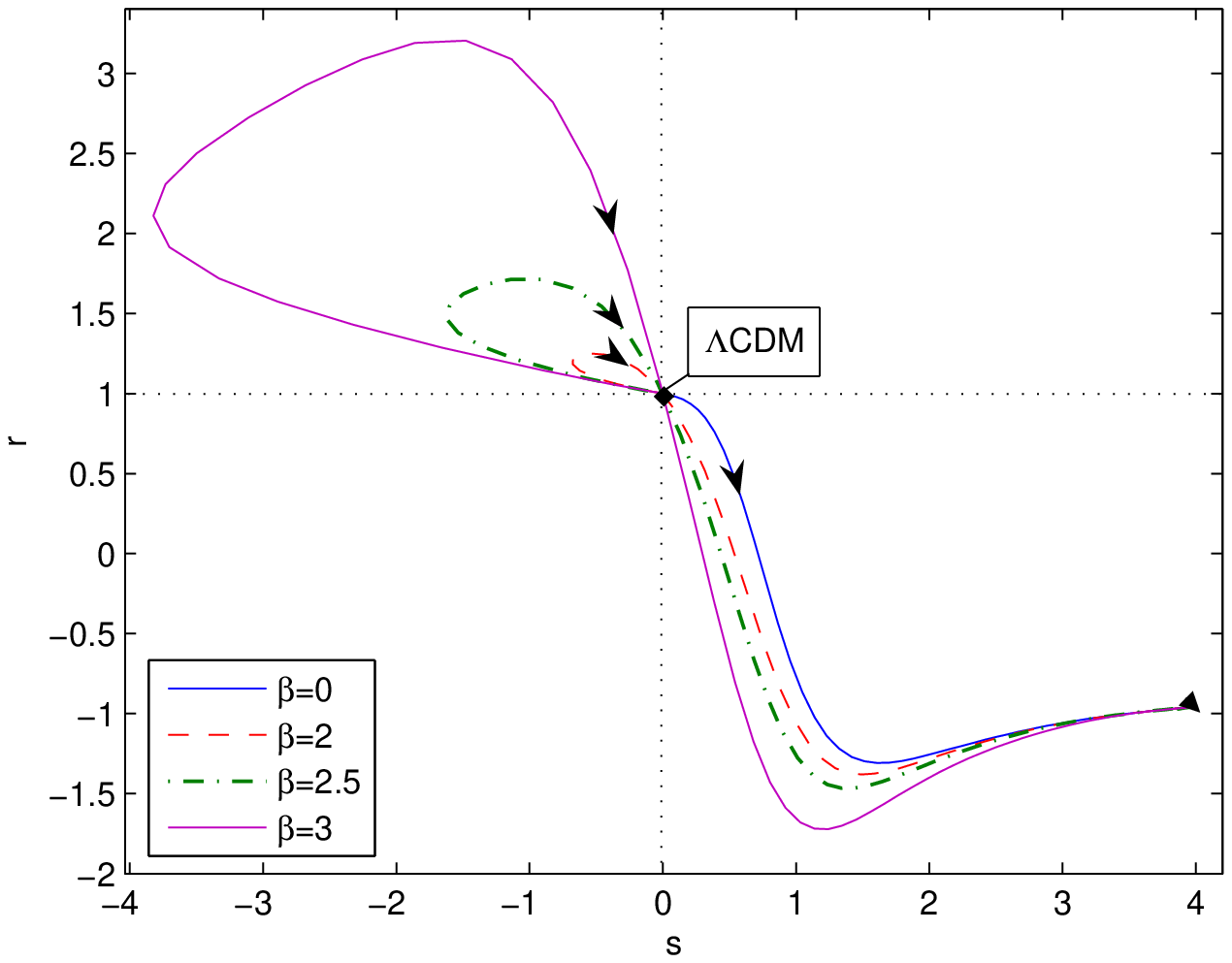}}
      \subfigure{\includegraphics[width=0.48\textwidth]{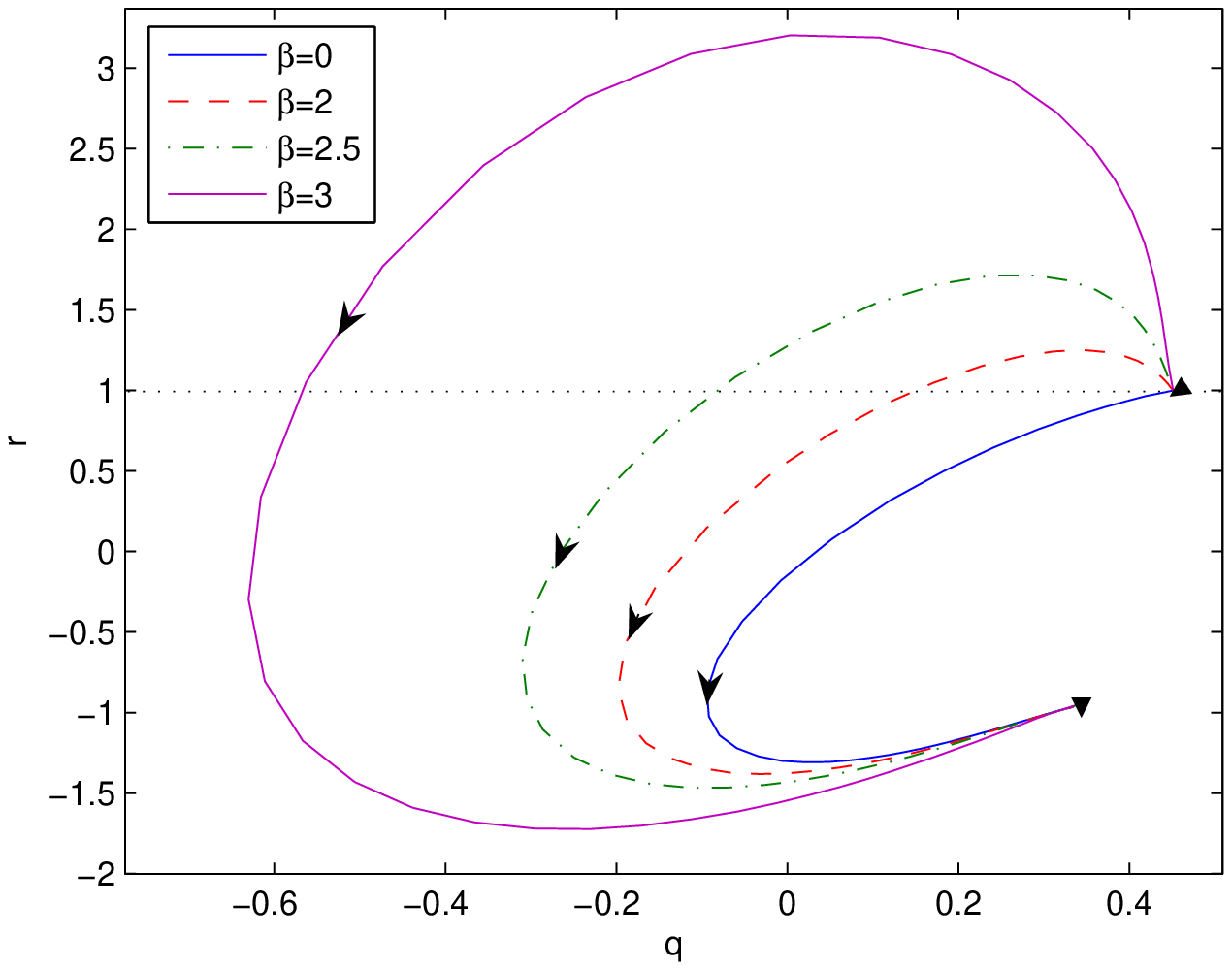}}}
\caption{The time evolution of the statefinder pairs $\{r,s\}$ and
$\{r,q\}$ is illustrated in left and right panel, respectively. Here
the fixed  model parameters are chosen to be $\gamma=1$ and
$\lambda_0=4$, whereas the thermal coupling constant $\beta$ takes
different values. For $\beta=0$ curve, $s$ monotonically increase
from $0$ to the value the attractor takes, while $r$ and $q$ first
decrease from the start point to a minimum value then increase to
the attractor. For $\beta\neq 0$ curves, $q$ has the similar
behavior of that of $\beta=0$ curve, whereas $r$ ($s$) first
increase (decrease) from $1$ ($0$)  to a maximum (minimum) value
then decrease (increase) and after passing through the $\Lambda$CDM
fixed point (in $r-s$ plane), they show a similar behavior of that
of $\beta=0$ curve.}\label{fig:l4g1}
\end{center}
\end{figure}

\begin{figure}
\begin{center}
\mbox{\subfigure{\includegraphics[width=0.48\textwidth]{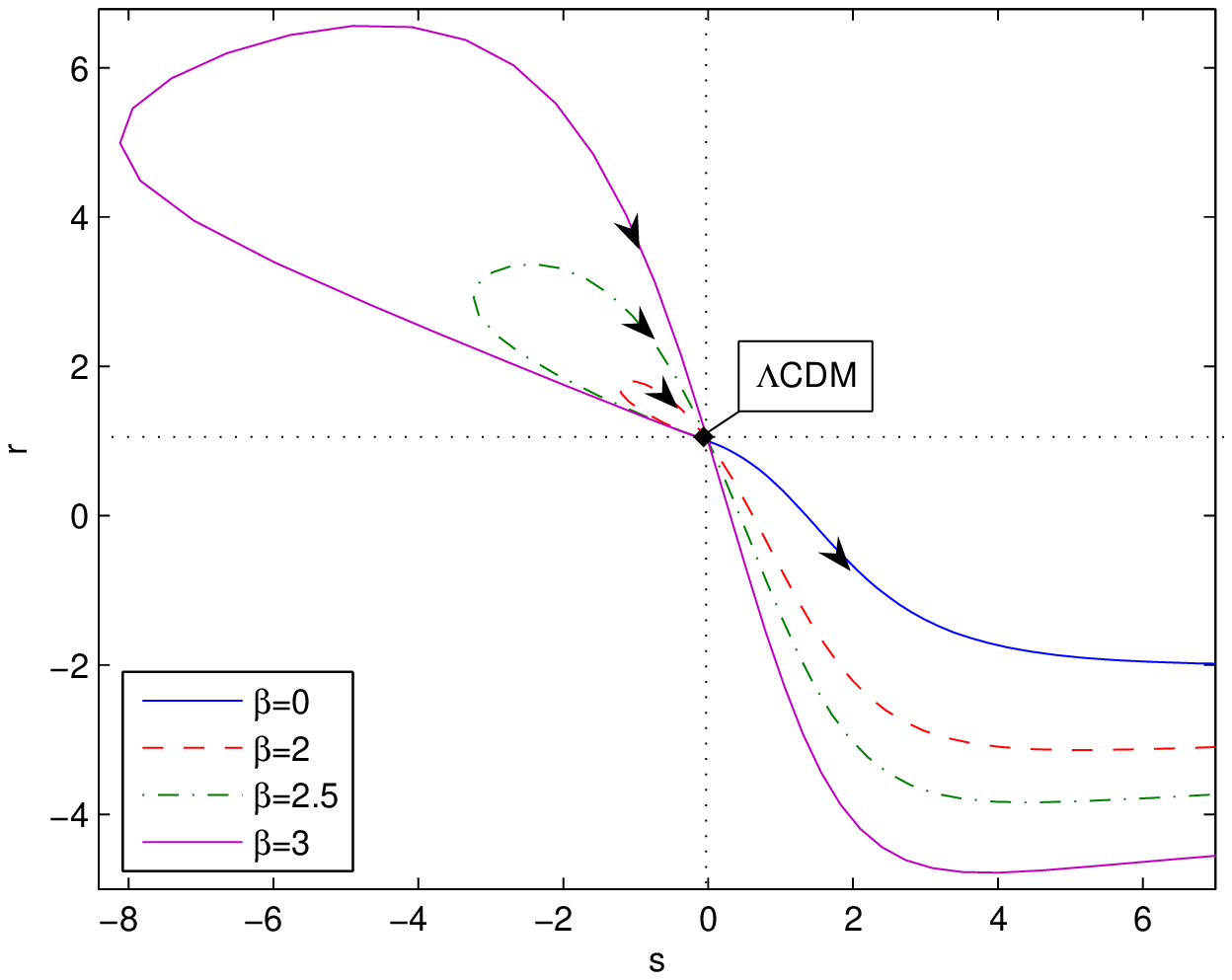}}
      \subfigure{\includegraphics[width=0.48\textwidth]{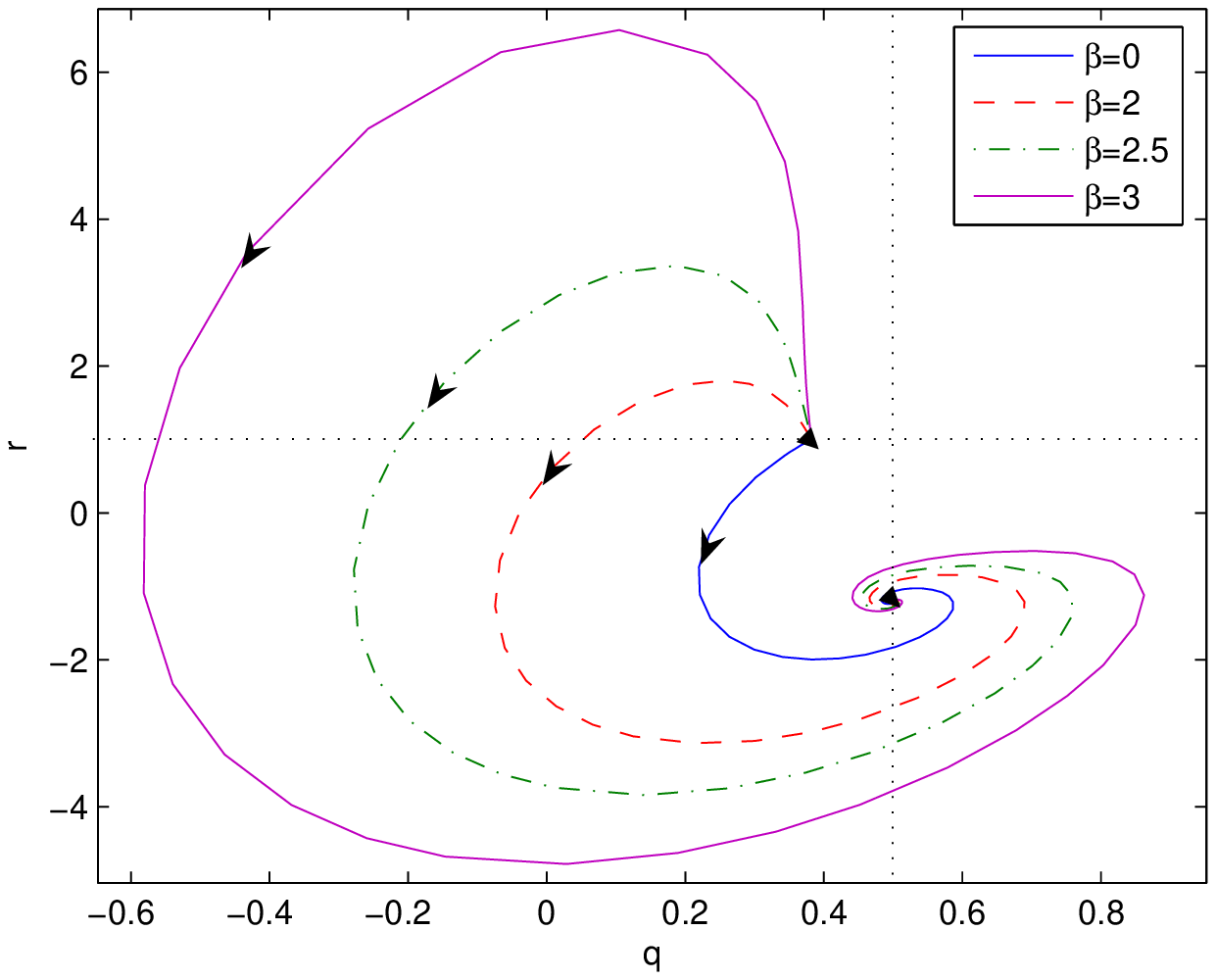}}}
\caption{The time evolution of the statefinder pairs $\{r,s\}$ and
$\{r,q\}$ is illustrated in left and right panel, respectively. Here
the fixed  model parameters are chosen to be $\gamma=1$ and
$\lambda_0=7$, while the thermal coupling constant $\beta$ takes
different values. Because $q$ approaches to and takes the value of
$1/2$ many times at the late time, but at the same time $r$ is
always less than unity, the value of $s$ will become infinity many
times. Therefore, the plot in the left panel shows only a piece of
the whole time evolution. }\label{fig:l7g1}
\end{center}
\end{figure}

We first fix the brayotropic parameter $\gamma$ and potential
parameter $\lambda_0$, and let the thermal coupling constant $\beta$
be adjustable. We show the time evolution of the statefinder pairs
$\{r,s\}$ (left panels) and $\{r,q\}$ (right panels) in Figure
\ref{fig:l1g1}, Figure \ref{fig:l4g1} and Figure \ref{fig:l7g1}
where the fixed parameters $\gamma=1$ and $\lambda_0=1,4,7$,
respectively. In these three figures, we choose an initial state
that the universe was ongoing a decreasing expansion and the kinetic
energy of the quintessence is negligible. The trajectories plotted
in Figure \ref{fig:l1g1} and Figure \ref{fig:l4g1} correspond to
case A. Obviously, they are all located in the second ($s<0,r>1$)
and/or fourth ($s>0,r<1$) quadrants in the $r-s$ plane. This is not
surprising because we have assumed that the equation of state of
quintessence field became negative after some past moment and this
makes the deceleration parameter $q<1/2$ from then on as is shown in
right panels of Figure \ref{fig:l1g1} and Figure \ref{fig:l4g1}.
From the definition of statefinder parameter $s$, Eq.(\ref{defr}),
it is easy to find that if $r<1$ then $s>0$, if $r>1$ then $s<0$ and
if $r=1$ and $q\neq 1/2$ then $s=0$ which is just corresponding to
the $\Lambda$CDM fixed point. It is worth pointing out that the
attractors in the two figures have different cosmological
implication: the attractor illustrated in Figure \ref{fig:l1g1}
means eternal acceleration of the universe, while the one in Figure
\ref{fig:l4g1} means the universe will decelerate in the future.
The trajectories plotted in Figure \ref{fig:l7g1} correspond to case
B. The accelerating rate  of expansion of the universe will be
determined only by the baryotropic matter.  We choose the
baryotropic parameter $\gamma=1$ in Figure \ref{fig:l7g1} which
means the universe will be decelerated in the future, as is shown in
the $r-q$ plane (right panel). From these three figures, we see that
the thermal coupling term  enhances the possible maximum value of
$r$ and makes it first increase to a maximum value ($>1$) then
decrease. This effect reflected in the $r-s$ plane is that some
parts of the $\beta\neq 0$ curves can be located in the second
quadrants ($r>1,s<0$) and these parts form closed loops. This
feature of the $r-s$ plane for the thermal interacting quintessence
model is distinguished from  those of ordinary quintessence and
other dark energy models.

\begin{figure}
\begin{center}
\includegraphics[width=\textwidth]{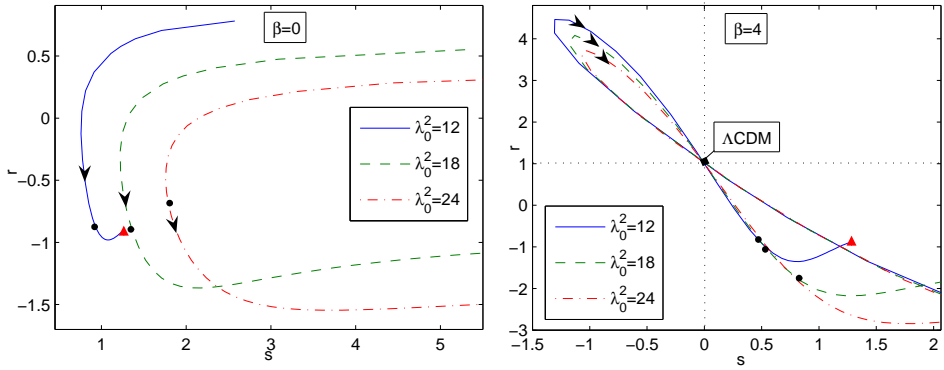}
\caption{The trajectories in the statefinder plane $\{r,s\}$ for the
model where we fix the baryotropic parameter $\gamma$ to be unity
and let the thermal coupling constant $\beta=0$ and $\beta=4$ in the
left and right panel, respectively.  The dark dots indicate the
current value of statefinder parameters. The triangle at the end of
the $\lambda_0^2=12$ curve denotes the location of the late-time
attractor. For the other two curves the attractor is located at
infinity.}\label{fig:rs-g1}
\end{center}
\end{figure}

\begin{figure}
\begin{center}
\includegraphics[width=\textwidth]{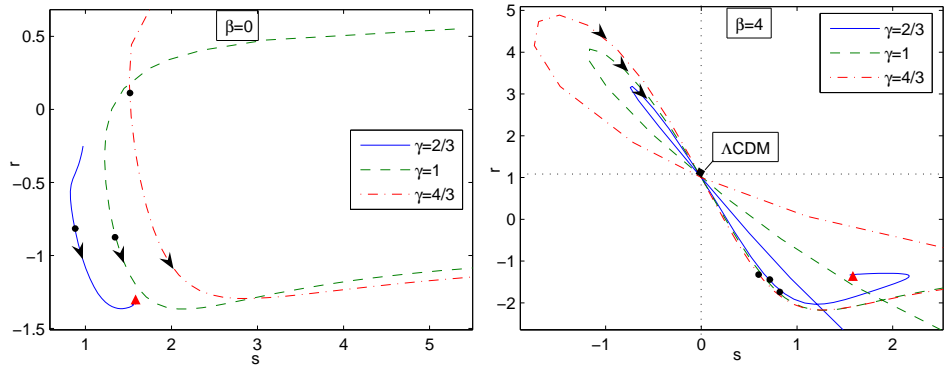}
\caption{The trajectories in the statefinder plane $\{r,s\}$ for the
model where we fix the baryotropic parameter $\lambda_0=\sqrt{18}$
and let the thermal coupling constant $\beta=0$ and $\beta=4$ in the
left and right panel, respectively.  The dark dots indicate the
current value of statefinder parameters. The triangle at the end of
the $\gamma=2/3$ curve denotes the location of the late-time
attractor. For the other two curves the attractor is located at
infinity. }\label{fig:rs-ll18}
\end{center}
\end{figure}

Next,  let us show the dependence of the time evolution of the
statefinder pair $\{r,s\}$ on the brayotropic parameter $\gamma$ and
potential parameter $\lambda_0$  in the absence or presence of the
thermal interaction. In Figure \ref{fig:rs-g1}, without loss of
generality the brayotropic parameter $\gamma$ is fixed to be unity
and the potential parameter is selected to satisfy the condition
$\lambda_0^2=12,18,24$, respectively. The left panel shows the
evolution for the case that thermal interaction is absent
($\beta=0$) and the right one for the case with thermal interaction
(we set $\beta=4$ here for example).Clearly, the effect that the
thermal coupling term exerts is the same as that pointed out
previously.  The $\lambda_0^2=12$ curve has a definite end point
which locates the late-time attractor but the others haven't. This
is not surprising. The parameter related to the $\lambda_0^2=12$
curve lies in the region A of the parameter space (see Figure
\ref{para}) and the location of the attractor is determined by Eqs.
(\ref{attractor1}) which take definite values, while the parameters
relevant to the other curves locate in the region B or the boundary
line between region A and B, so the locations of the corresponding
attractor is determined by Eqs. (\ref{attractor2}) which are
infinite for the parameters selected. The situation is similar in
Figure \ref{fig:rs-ll18}, where we fix the value of $\lambda$ and
select three different representative values for $\gamma$. In
addition, the dark dots marked on the curves in the two figures
represent the present values of the statefinder parameters
$\{r_0,s_0\}$. It should be noted that the true values of
$\{r_0,s_0\}$ of the universe should be determined in a
model-independent way, which we can only place our hope on the
future experiments. In principle, if this is achieved, the
statefinder parameters can be viewed as a discriminator for testing
various cosmological models.

\section{Conclusions}
We have investigated the time evolution of quintessence field with
an exponential potential and with or without thermal coupling to the
background matter in a flat FRW universe from the statefinder
viewpoint in this paper. The statefinder diagnostic have been
performed. It is found that that the thermal coupling term does not
affect the location of the late-time attractor which is determined
only by the potential of the quintessence field and the background
baryotropic matter, but exert an influence on the evolution of the
statefinder parameters. The most notable feature  for the thermal
coupling model which is distinguished from the other dark energy
models is that  the curve with thermal coupling can form a closed
loop in the second quadrant ($r>1,s<0$) in the $r-s$ plane. What we
expect now is that more and more accurate observational data could
be offered to determine the model parameters more and more
precisely, rule out some models and consequently shed light on the
essence of dark energy.

\section*{Acknowledgments}
This work is supported in part by National Natural Science
Foundation of China under Grant No. 10503002 and Shanghai Commission
of Science and technology under Grant No. 06QA14039.

\end{document}